\documentclass[12pt]{article}
\usepackage{amssymb}
\usepackage{latexsym}
\usepackage{amsmath}
\usepackage{graphicx}

\def\vp{\varphi}
\def\to{\tilde{\omega}}
\def\bo{\bar{\omega}}
\def\bm{\bar{m}}
\title{\LARGE{Fermionic quasinormal modes for two-dimensional Ho\v{r}ava-Lifshitz black holes }}
\author{M. M. Stetsko\footnote{E-mail: mstetsko@gmail.com}\
\\
  {\small Department for Theoretical Physics, Ivan Franko National University of Lviv,}\\
{\small 12 Drahomanov Str., Lviv, UA-79005, Ukraine
         }}

\begin{document}
\maketitle

\abstract{To obtain fermionic quasinormal modes, the Dirac equation for two types of black holes is investigated. For the first type of black hole, the quasinormal modes have continuous spectrum with negative imaginary part that provides the stability of black hole geometry. For the second type of the black hole, the quasinormal modes have discrete spectrum and are completely imaginary. This type of the black hole appears to be stable for arbitrary masses of fermion field perturbations.}

\section{Introduction}
The investigation of gravitational perturbations of the Schwarzschild geometry  started several decades ago \cite{Regge_PR1957, Zerilli_PRD1970,Zerilli_PRL1970}.  That idea was applied for examination of perturbations of other types of black holes caused by the fields of different nature, for example scalar field or Dirac field.  All those works gave birth to the method which is known nowadays as the quasinormal modes  method  (QNM)\cite{Nollert_CQG99,Berti_CQG2009, Konoplya_RMP2011}. This method allows one to get important information about stability of black holes under the influence of perturbations of different types which evolve in the exterior region of the black holes. We also note that in most cases the influence of the external fields is considered perturbatively and the backreaction of the field on the black hole\rq{}s metric is not taken into account.  The quasinormal modes and their quasinormal frequencies are useful for different branches of investigations in general realtivity. In particular, QN modes are important  in gauge-string duality theories (AdS-CFT) \cite{aharony_PhysRept2000} because they define relaxation times of dual field theories \cite{Horowitz}.  The relation between QN modes and retarded correlators of dual field theories was also established \cite{Son_JHEP2002, Starinets_PRD02, Nunez_PRD03, Kovtun_PRD05}. Another  possibility is due to Hod\rq{}s conjecture  about quantization of black hole\rq{}s area \cite{Hod_PRL98, Dreyer_PRL2003, Maggiore_PRL2008}.  The connection between QN modes and Hawking radiation is also considered \cite{Konoplya_PRD2004, Kiefer_CQG2004}.  The progress in experimental astrophysics and discovery of gravitational waves open a new perspective for for application of QNM method for estimation of different parameters of compact sources of graviational field or verifictation some conjectures of general relativity\cite{Berti_CQG2009}.

Another area of active research is related to the two different disciplines, namely quantum mechanics and general realtivity. A well known and still open problem is reconcilation of principles of these theories. It might give some hints about underlying theory of quantum gravity. For example nonrenormalizability is a crucial problem when one tries to quantize general relativity in the way possible for other gauge fields.  To overcome this difficulty, it was supposed that the general relativity should be treated as an effective theory and in order to have the graviation theory suitable for quantization the principles of general relativity should be elaborated. One of the approaches that leads to power-countable UV-renormalizability is so called Ho\v{r}ava-Lifshitz theory \cite{Horava_PRD09, Horava_JHEP09, Horava_PRL09}. General Relativity can be recovered as an infrared limit of the Ho\v{r}ava-Lifshitz theory.  Because of its attracive and promising features, the Ho\v{r}ava-Lifshitz approach has gained considerable interest in recent years. In particular, black hole solutions were found and their properties were investigated \cite{Kehagias_PLB09, Park_JHEP09, Cai_PRD09, Kiritsis_JHEP, Myung_PLB09, Cai_PLB09, Myung_EPJC10, Koutsoumbas_PRD10, Chen_PLB10}. Quasinormal modes for HL black holes were studied in the works \cite{Konoplya_PLB09, Varghese_MPLA2011, Varghese_GRG2011, Becar_IJMPD13, Cruz_arxiv_2015}.

The examination of fields evolution in a background of lower dimensional black holes is an interesting and important problem. Firstly, because of simplicity of those problems in comparison with higher dimensional cases anlytical computations can be made and, as a result, in many kinds of black holes exact QN frequencies can be calculated. The second important point  is the fact that lower dimensional black holes and the fields evolving in their backgrounds gave some hints or reveal some aspects of highr dimensional cases.

Our paper is organised as follows: in the section 2 we briefly review some $1+1$ dimensional black hole solutions in HL gravity. In the section 3 the Dirac equations for fermion fields in specific black holes background are written. In the section 4 we investigate fermionic QNMs  for chosen BH metrics. The last section contains some concluding remarks.

\section{$1+1$ dimensional black holes solutions in Ho\v{r}ava theory}
$1+1$ dimensional black holes in Ho\v{r}ava gravity were considered in the paper \cite{Bazeia_PRD2015}. The starting point is two dimensional action integral which in case of Ho\v{r}ava-Lifshitz (HL) gravity takes the form:
\begin{equation}
S=\frac{M^2_{Pl}}{2}\int dtdx\left(-\frac{1}{2}\eta N^2a^2+\alpha N^2{\vp\rq{}}^2-V(\vp)\right),
\end{equation}
and here $\alpha$, $\eta$ are constants and $a=N\rq{}/N=(\ln N)\rq{}$. 

The black hole solutions in two dimensional HL gravity is described by a shift function which can be represented as follows:
\begin{equation}
N^2(x)=2C_2+\frac{A}{\eta}x^2-2C_1x+\frac{B}{\eta x} +\frac{C}{3\eta x^2} 
\end{equation}
where $A, B, C, C_1, C_2$ are some constants. The scalar potential can be written in the form:
\begin{equation}
\vp(x)=\frac{1}{2}\ln\left(2C_2+\frac{A}{\eta}x^2-2C_1x+\frac{B}{\eta x} +\frac{C}{3\eta x^2}\right)
\end{equation}
Written expression for the shift function as well as the scalar potential are quite general and we consider some particular cases, taking specific values for the mentioned above constants. We describe them in the following sequence.
\begin{itemize}
\item The first case: for $C_1=-M$, $C_2=-\frac{1}{2}$ and $A=B=C=0$ we have $V_{\vp}(\vp)=0$ (or $V(\vp)=const$) and one arrives at the solution:
\begin{equation}\label{metric_1}
ds^2=-(2M|x|-1)dt^2+\frac{1}{(2M|x|-1)}dt^2
\end{equation}
It should be noted that similar solution was obtained in the context of the ordinary $1+1$ dimensional gravity \cite{Mann90}.
\item The second case: the constants are chosen in the following way: $A=\Lambda$, $B=C=0$, $C_1=-M$ and $C_2=-\epsilon/2$. For this case we have $V_{\vp}(\vp)=\Lambda$ which leads to linear dependence for the scalar potential $V(\vp)=\Lambda\vp$. The solution takes the form:
\begin{equation}\label{metric_2}
ds^2=-\left(\left(\Lambda/\eta\right)^2x^2+2Mx-\epsilon\right)dt^2+\frac{1}{\left(\left(\Lambda/\eta\right)^2x^2+2Mx-\epsilon\right)}dx^2
\end{equation} 
The latter metric can be rewritten in a bit different form after some kind of transformation of coordinates \cite{Cruz_arxiv_2015}:
\begin{equation}
u=\sqrt{\frac{\Lambda}{\eta}}x+\sqrt{\frac{\eta}{\Lambda}}M.
\end{equation}
Having used written above transformation we arrive at a new form of the metric (\ref{metric_2}):
\begin{equation}\label{metric_2m}
ds^2=-(u^2-u^2_+)dt^2+\frac{l^2}{(u^2-u^2_+)}du^2
\end{equation}
and here $u_+=\sqrt{(\eta/\Lambda)M^2+\epsilon}$ and $l=\sqrt[4]{\Lambda/\eta}$. It is worth noting that in the new coordinate system the horizons of the black hole are located at the point: $u=u_+$.
\item The third case: so called Schwarzschild-like solution. In this case 
one imposes that $A=C=C_1=0$, $B=-2M$, $C_2=1/2$ and $\eta=1$.  As a result the metric would look as follows: 
\begin{equation}
N^2(x)=1-\frac{2M}{x}, \quad \vp(x)=\frac{1}{2}\ln\left(1-\frac{2M}{x}\right)
\end{equation}
We also note that in this case the potential can be written in explicit form \cite{Bazeia_PRD2015}. So the metric takes Schwarzshild-like form:
\begin{equation}
ds^2=-\left(1-\frac{2M}{x}\right)dt^2+\frac{1}{\left(1-\frac{2M}{x}\right)}dx^2
\end{equation}
\item The fourth case is so called Reissner-Nordstr\"{o}m-like case. The constants should be chosen as follows: $A=C_1=0$, $B=-2M$, $C=3Q^2$ and $C_2=1/2$. So we obtain
\begin{equation}
N^2(x)=1-\frac{2M}{x}+\frac{Q^2}{x^4}, \quad \vp(x)=\frac{1}{2}\ln\left(1-\frac{2M}{x}+\frac{Q^2}{x^4}\right)
\end{equation} 
As a result the metric takes the Reissner-Nordstr\"{o}m-like form
\begin{equation}
ds^2=-\left(1-\frac{2M}{x}+\frac{Q^2}{x^2}\right)dt^2+\frac{1}{\left(1-\frac{2M}{x}+\frac{Q^2}{x^2}\right)}dx^2
\end{equation}
We note that in contrast to the previous cases here, it is not possible to find explicit form for the scalar potential.
\end{itemize}
\section{Dirac equation}
Fermionic perturbations in the background of two dimensional black holes is governed by the Dirac equation. Supposing that the fermionic fields  are chargeless, we can write:
\begin{equation}
\left(\gamma^{\mu}\nabla_{\mu}+m\right)\psi=0,
\end{equation}
where $m$ is the mass of fermionic field $\psi$. The covariant derivative is defined as follows:
\begin{equation}
\nabla_{\mu}=\partial_{\mu}+\frac{1}{2}\omega^{AB}_{\mu}J_{AB}
\end{equation}
where $J_{AB}=\frac{1}{4}[\gamma_{A},\gamma_{B}]$  are the Lorentz group generators and $\omega^{AB}_{\mu}$ denotes components of spin connection. Gamma matrices in a curved space take the form $\gamma^{\mu}=e^{\mu}_A\gamma^A$ where $e^{\mu}_A$ are the diad components and $\gamma^A$ are gamma matrices for flat space-time. To obtain the spin connection, the Cartan structure equation should be utilized
\begin{equation}\label{cartan}
de^{A}+{\omega^{A}}_{B}\wedge e^{B}=0
\end{equation}
The connection is supposed to be torsionless. It can be easily verified that for all the cases we have mentioned above the only nonzero component of spin connection will be the component $\omega^{01}$. We will consider the first two cases and the others will be investigated elsewhere.

\subsection{Dirac equation for the first kind of the metric}
In this case the diad takes the form as follows:
\begin{equation}
e^0=\sqrt{2Mx-1}dt, \quad e^1=\frac{1}{\sqrt{2Mx-1}}dx
\end{equation} 
Having used the equation (\ref{cartan}), we obtain the spin connection:
\begin{equation}
{\omega^{0}}_{1}=\frac{M}{\sqrt{2Mx-1}}e^0
\end{equation}
For  gamma matrices $\gamma^{A}$ (lorentzian) the following representation will be used:
\begin{equation}
\gamma^0=i\sigma^2, \quad \gamma^1=\sigma^1,
\end{equation}
 where $\sigma^i$ are the Pauli matrices.
 In the curvilinear coordinates, the gamma matrices look as follows:
 \begin{equation}
 \gamma^t=\frac{1}{\sqrt{2Mx-1}}\gamma^0, \quad \gamma^x=\sqrt{2Mx-1}\gamma^1
 \end{equation}
 Now the Dirac equation for the metric (\ref{metric_1}) can be written:
 \begin{equation}
 \left(\frac{i\sigma^2}{\sqrt{2Mx-1}}\left(\partial_t-\frac{M}{2}\sigma^3\right)+\sigma^1\sqrt{2Mx-1}\partial_x+m\right)\psi=0
\end{equation} We suppose that the solution of the written equation can be represented in the form:
\begin{eqnarray}\label{wave_funct_1}
\psi(t,x)=\frac{1}{\sqrt[4]{2Mx-1}}e^{-i\omega t}
\begin{pmatrix}
\psi_1 \\
\psi_2 \\
\end{pmatrix}
\end{eqnarray}
The system of equations for the components of the spinor part of the wavefunction takes the form:
\begin{eqnarray}\label{coupled_dirac_1}
\nonumber\frac{-i\omega}{\sqrt{2Mx-1}}\psi_2+\sqrt{2Mx-1}\partial_x\psi_2+m\psi_1=0,\\
\frac{i\omega}{\sqrt{2Mx-1}}\psi_1+\sqrt{2Mx-1}\partial_x\psi_1+m\psi_2=0,
\end{eqnarray}
The system of the equations, we have obtained, can be decoupled and we write the equation for one component of the wavefunction. For example for the function $\psi_1$:
\begin{equation}\label{dirac_eq_1_1}
(2Mx-1)\partial^2_x\psi_1+M\partial_x\psi_1+\left(\frac{\omega^2-i\omega M}{2Mx-1}-m^2\right)\psi_1=0
\end{equation}
\subsection{Dirac equation for the second kind of the metric}
Now we consider Dirac equation for the second kind of the metric that we have written (\ref{metric_2}). The transformed form of the metric represented by the relation (\ref{metric_2m}) will be used here. The diad field for this metric takes the form:
\begin{equation}
e^0=\sqrt{u^2-u^2_+}dt, \quad e^1=\frac{l du}{\sqrt{u^2-u^2_+}}
\end{equation}
Having used the Cartan structure equation (\ref{cartan}), we obtain following expression for the spin connection form:
\begin{equation}
{\omega^0}_1=\frac{u}{l\sqrt{u^2-u^2_+}}e^0.
\end{equation}
Finally, the Dirac equation for the background metric (\ref{metric_2m}) can be written as follows:
\begin{equation}\label{dirac_2d}
\left(\frac{l}{\sqrt{u^2-u^2_+}}i\sigma^2\left(\partial_t-\frac{u}{2l}\sigma^3\right)+\sqrt{u^2-u^2_+}\sigma^1\partial_u\right)\psi+\bar{m}\psi=0,
\end{equation}   
where $\bar{m}=lm$.
Similarly to the previous case, in order to simplify the procedure of the solution of the written above Dirac equation, we assume that the spinor wave function takes the following form:
\begin{eqnarray}\label{wave_funct_2}
\psi=\frac{e^{-i\omega t}}{\sqrt[4]{u^2-u^2_+}}
\begin{pmatrix}
\psi_1 \\
\psi_2 \\
\end{pmatrix}
\end{eqnarray} 
Having substituted the wave function (\ref{wave_funct_2}) into the equation (\ref{dirac_2d}), and after little algebra, we obtain the system of equations for the componets of spinor wave function $\psi_1$ and $\psi_2$:
\begin{eqnarray}
\left(-\frac{il\omega}{\sqrt{u^2-u^2_+}}+\sqrt{u^2-u^2_+}\partial_u\right)\psi_2+\bar{m}\psi_1=0,\\
\left(\frac{il\omega}{\sqrt{u^2-u^2_+}}+\sqrt{u^2-u^2_+}\partial_u\right)\psi_1+\bar{m}\psi_2=0\label{dir_2c}
\end{eqnarray}

The system of equations can be easily decoupled and, as a result, the equation for any component of the spinor wave function (\ref{wave_funct_2}) can be obtained. We write the equation for the component $\psi_1$ :
\begin{equation}\label{eq_qnm_2}
\left((u^2-u^2_+)\partial^2_u+u\partial_u+\frac{\bar{\omega}^2-i\bar{\omega}u}{u^2-u^2_+}-\bar{m}^2\right)\psi_1=0
\end{equation}
and here $\bar{\omega}=l\omega$. 
\section{Quasinormal modes}
In this section, the Dirac equations for to cases of  metric will be examined again separately. We will study quasinormal modes and then compare the results. We also remark that quasinormal modes for scalar perturbations in the same black holes background were considered in the article \cite{Cruz_arxiv_2015}.    
\subsection{Quasinormal modes for the metric of the first kind}
In this subsection, we will consider equation (\ref{dirac_eq_1_1}) and investigate quasinormal modes for it. Firstly, we make a transformation of coordinates and introduce a new one instead of coordianate $x$ by the following relation:
\begin{equation}
z=\frac{m}{M}\sqrt{2Mx-1}
\end{equation}
The equation (\ref{dirac_eq_1_1}) can be rewritten in the form:
\begin{equation}
\partial^2_z\psi_1+\frac{\to^2-i\to}{z^2}\psi_1-\psi_1=0
\end{equation}
and here $\to=\omega/M$. We suppose that the wave function $\psi_1$ of the latter equation takes the form:
\begin{equation}
\psi_1=\sqrt{z}F(z)
\end{equation}
Having performed that transformation, we will arrive at the modified Bessel equation:
\begin{equation}\label{mod_bessel}
z^2F^{\prime\prime}+zF^{\prime}-(\nu^2+z^2)F=0.
\end{equation}
where $\nu=1/2+i\tilde{\omega}$.  The solutions of the latter equation are well known modified Bessel functions \cite{abramowitz} and can be used for analysis of quasinormal modes:
\begin{equation}
F(z)=AI_{\nu}(z)+BK_{\nu}(z)
\end{equation}
To obtain quasinormal modes we have to impose boundary conditions on the solutions of the wave equation. It is known that in the vicinity of the horizon the solution of the corresponding wave equation should behave as an ingoing wave. The behaviour at the infinity depends on the background metric. In case of an asymptotically flat geometry the solution of the wave equation should behave as an outgoing wave. In our case the geometry is not asympotically flat, the metric function diverges at the infinity. As a consequence, we impose that the wave function of our equation  (\ref{mod_bessel}) should tend to zero at the infinity.  The function $I_{\nu}(z)$ is divergent  when $z\rightarrow +\infty$ for arbitrary value of the parameter $\nu$ whereas $K_{\nu}(z)$ has nondivergent behaviour. to get rid of the term divergent at the infinity we put $A=0$. As a result, the solution which fulfil the boundary condition at the infinity takes the form:
\begin{equation}\label{bess_K}
F(z)=BK_{\nu}(z)
\end{equation}
In the vicinity of the horizon we use asymptotic formula for the Bessel function $K_{\nu}(z)$ \cite{abramowitz}:
\begin{equation}\label{bess_decomp}
K_{\nu}(z)\simeq \frac{1}{2}\Gamma(\nu)\left(\frac{z}{2}\right)^{-\nu}
\end{equation}
The written above decomposition is valid when ${\rm Re}(\nu)>0$.  The latter condition leads to the restriction on the imaginary part of the frequency, namely ${\rm Im}(\omega)=\omega_I<M/2$. For the upper component of the Dirac wavefunction we obtain:
\begin{equation}
\psi_{up}\sim e^{-i\omega t}z^{-i\tilde{\omega}}=e^{-i\omega(t+1/M\ln{z})}.
\end{equation}
As one can see in the domain close to the horizon the upper component of the Dirac wavefunction behaves as an ingoing wave as it should be for the quasinormal modes. To make them stable one should impose that  the imaginary part of the quasinormal modes negative. When ${\rm Re}(\nu)<0$ we use the well-known relation for the Bessel function $K_{\nu}(z)$:
\begin{equation}\label{bess_pr}
K_{\nu}(z)=K_{-\nu}(z)
\end{equation}
And in the domain close to the horizon we also use  approximate relation:
\begin{equation}
K_{-\nu}(z)\simeq \frac{1}{2}\Gamma(-\nu)\left(\frac{z}{2}\right)^{\nu}
\end{equation}
It was shown that in this case near the horizon the upper component for the Dirac wavefunction behaves as an outgoing  wave so it does not satisfy the boundary condition for the quasinormal modes \cite{Estrada-Jimenez_2013}. To analyze the behaviour of the lower component we use the relation (\ref{coupled_dirac_1}) and write:
 \begin{equation}\label{coup_2}
 \psi_{2}=-\left(\frac{i\tilde{\omega}}{z}+\frac{\partial}{\partial z}\right)\psi_1=-z^{-i\tilde{\omega}}\frac{\partial}{\partial z}z^{i\tilde{\omega}}\psi_1=-z^{1/2-\nu}\frac{\partial}{\partial z}z^{\nu}F(z)
 \end{equation} 
Having substituted the solution (\ref{bess_K}) into the written above relation and taking into account relations  for derivatives of the Bessel function we obtain:
\begin{equation}\label{lower_comp}
 \psi_{2}=-Bz^{1/2}K_{\nu-1}(z)
\end{equation}
The lower component (\ref{lower_comp}) of the Dirac wave function similarly to the upper one tends to zero at the ininity. To examine the behaviour in the domain close to the horizon we again use the relation (\ref{bess_decomp}) and write:
\begin{equation}
 \psi_{2}\simeq-\frac{B}{2}\Gamma(\nu-1)z^{1/2}\left(\frac{z}{2}\right)^{-\nu+1}
\end{equation}
The written relation takes place when ${\rm Re}(\nu-1)>0$ which is equivalent to the condition $\omega_I<-\frac{M}{2}$. Similarly to the upper component the lower component also behaves as an ingoing wave in the vicinity of the horizon:
\begin{equation}
\psi_{down}\simeq ze^{-i\omega(t+1/M\ln{z})}.
\end{equation}
When ${\rm Re}(\nu-1)<0$ the relation (\ref{bess_pr}) can be used again, but it can be shown that in this case one obtains outgoing waves close to the horizon which does not satisfy boundary condition at the infinity. We can conclude that both components of the Dirac wavefunction might satisfy the necessary conditions for the quasinormal modes and to make them stable we have to impose $\omega_I<-M/2$. We note that the similar conclusion was made in the paper \cite{Estrada-Jimenez_2013}. We also remark that for integer $\nu$ one of the solutions of the equation (\ref{mod_bessel}) namely the function $K_n(z)$ can be introduced as limit of the function $K_{\nu}(z)$ when $\nu\rightarrow n$. The behaviour of the solution $K_n(z)$  at the infinity and at the horizon is similar to the case of noninteger $\nu$ and it means that they also satisfy the mentioned above conditions for the quasinormal modes, but there is no specific requirement that might distinguish integer values of the parameter $\nu$ from the noninteger. 

In the case of scalar particles the behaviour of particle flux at the infinity was also analyzed \cite{Cruz_arxiv_2015}. Taking into account the definition of the flux for the Dirac particles we can write:
\begin{equation}\label{flux_1}
{\cal F}=\sqrt{-g}\bar{\psi}\gamma^r\psi
\end{equation}
and here $\gamma^r=e^r_1\gamma^1$, $\bar{\psi}=\psi^{\dagger}\gamma^0$, $\sqrt{-g}=1$. As a consequence we obtain:
\begin{equation}\label{flux}
{\cal F}=|\psi_1|^2-|\psi_2|^2
\end{equation}
The behaviour of the Dirac flux at the infinity is completely defined  by the corresponding behaviour of the upper and lower components of the Dirac wavefunction. Because both of them tend to zero at the infinity the Dirac flux vansihes at the infinity. It is worth being emphasized  that in case of scalar particles the vanishing behaviour of the flux might be provided when one imposes that the scalar field vanishes at the infinity (Dirichlet condition) or its derivative disappears at the infinity (Neumann condition).  For Dirac particles there is no specific requirement for the derivatives of upper or lower components of the Dirac wavefunction but both of them are connected through the relations (\ref{coupled_dirac_1}) or (\ref{coup_2}) where derivatives from components are present, so in some way boundary conditions on the upper and lower components are equivalent to imposing both Neumann and Dirichlet conditions.
 
 \subsection{Quasinormal modes for the second kind of the metric}
Similarly to the previous case, we have to  solve the equation (\ref{eq_qnm_2}). The equation can be rewritten in the form of a standard hypergeometric equation. To simplify the calculations, we perform a transformation of coordinates defined by the following relation:
\begin{equation}\label{transf_u_z}
z=\frac{u-u_+}{u+u_+}
\end{equation}
It can be verified easily that the domain of variation of the variable $z$ is the interval: $-1\leqslant z\leqslant 1$ and since we consider the motion of the particle outside the black hole our domain will be as follows: $0\leqslant z\leqslant 1$.
Having used the transformation (\ref{transf_u_z}) we rewrite the equation (\ref{eq_qnm_2}) in the form:
\begin{equation}
z(1-z)^2\psi^{\rq{}\rq{}}_1+(1-z)\left(\frac{1}{2}-\frac{3}{2}z\right)\psi^{\rq{}}_1+\left(\frac{\bar{\omega}^2(1-z)^2}{z}-\frac{i\bar{\omega}}{2z}(1-z)(1+z)-\bar{m}^2\right)\psi_1=0
\end{equation}
We suppose that the wave function $\psi_1$ can be represented in the form:
\begin{equation}\label{psi_1}
\psi_1=z^{\alpha}(1-z)^{\beta}F(z)
\end{equation}
As a consquence, a hypergeometric equation for the function $F(z)$ can be written:
\begin{equation}\label{hypergeo}
z(1-z)F^{\rq{}\rq{}}(z)+(c-(a+b+1)z)F^{\rq{}}(z)-abF(z)=0,
\end{equation}
where the coefficients $a$, $b$ and $c$ are given by the relations:
\begin{eqnarray}\label{comb_abc}
a+b=2(\alpha+\beta)+\frac{1}{2},\\
ab=(\alpha+\beta)^2+\frac{1}{2}(\alpha+\beta)+\bo^2+\frac{i}{2}\bo,\\
c=2\alpha+\frac{1}{2}.
\end{eqnarray}
The parameters should satisfy the system of equations:
\begin{eqnarray}
\alpha^2-\frac{\alpha}{2}+\bo^2-\frac{i}{2}\bo=0,\\
\beta^2-\alpha^2+\frac{\alpha}{2}-\bo^2+\frac{i}{2}\bo-\bm^2=0.
\end{eqnarray}
The latter system of equations can be solved easily and we obtain:
\begin{eqnarray}
\alpha_1=i\bo+\frac{1}{2}, \quad \alpha_2=-i\bo,\\
\beta_1=\bm, \quad \beta_2=-\bm
\end{eqnarray}
We note that any combination of $\alpha$ and $\beta$ can be chosen and substituted into the system of equations (\ref{comb_abc}).  We consider different combinations of the parameters and analyse the solution we arrive at.  Let us start from the combination of parameters $\alpha_1=i\bo+\frac{1}{2}$ and $\beta_1=\bm$. As a result from the system (\ref{comb_abc}) we obtain: 
\begin{equation}\label{comb_1}
a=\bm+\frac{1}{2}+2i\bo, \quad b=\bm+1  ,\quad c=2i\bo+\frac{3}{2}. 
\end{equation}
It is known that the general solution of the hypergeometric equation (\ref{hypergeo}) can be represented in the form \cite{abramowitz}:
\begin{equation}\label{gen_sol}
F=A{_{2}F_{1}}(a,b,c;z)+Bz^{1-c}{_{2}F_{1}}(a-c+1,b-c+1,2-c;z).
\end{equation}
The combiantion of parameters $a$, $b$, $c$ (\ref{comb_1}) and the written above general solution of the  hypergeometric equation (\ref{gen_sol})  immediately lead us to the solution of the hypergeometric equation (\ref{hypergeo}) which  takes the form:
\begin{equation}
F=A{_{2}F_{1}}\left(\bm+\frac{1}{2}+2i\bo,\bm+1,2i\bo+\frac{3}{2};z\right)+Bz^{-1/2-2i\bo}{_{2}F_{1}}\left(\bm, \bm+\frac{1}{2}-2i\bo,\frac{1}{2}-2i\bo;z\right)
\end{equation}
Now we choose another combination of the parameters: $\alpha_1=i\bo+1/2$ and $\beta_2=-\bm$. As a consequence we obtain:
\begin{equation}
a=\frac{1}{2}-\bm+2i\bo, \quad b=1-\bm  ,\quad c=2i\bo+\frac{3}{2}.
\end{equation}
The general solution for that combination of parameters takes the form:
\begin{equation}
F=A{_{2}F_{1}}\left(\frac{1}{2}-\bm+2i\bo, 1-\bm,2i\bo+\frac{3}{2};z\right)+Bz^{-1/2-2i\bo}{_{2}F_{1}}\left(-\bm, \frac{1}{2}-\bm-2i\bo,\frac{1}{2}-2i\bo;z\right)
\end{equation}
The third variant for the parameters $\alpha$ and $\beta$ can be taken as follows: $\alpha_2=-i\bo$ and $\beta_1=\bm$. For the chosen combination we obtain:
\begin{equation}
a=\bm,\quad b=\bm+\frac{1}{2}-2i\bo, \quad c=\frac{1}{2}-2i\bo
\end{equation}
So, we write the general solution of the hypergeometric equation  in the form:
\begin{equation}
F=A{_{2}F_{1}}\left(\bm, \bm+\frac{1}{2}-2i\bo,\frac{1}{2}-2i\bo;z\right)+Bz^{1/2+2i\bo}{_{2}F_{1}}\left(\bm+\frac{1}{2}+2i\bo, \bm+1,\frac{3}{2}+2i\bo;z\right)
\end{equation}
The last combination of the parametres $\alpha$ and $\beta$ that we can choose is $\alpha_2=-i\bo$ and $\beta_2=-\bm$. Taking this combination into consideration, we obtain:
\begin{equation}
a=-\bm, \quad b=\frac{1}{2}-\bm-2i\bo, \quad c=\frac{1}{2}-2i\bo
\end{equation}
The corresponding general solution of the hypergeometric equation for the obtained above parameters $a$, $b$ and $c$ can be represented in the form:
\begin{equation}
F=A{_{2}F_{1}}\left(-\bm, \frac{1}{2}-\bm-2i\bo, \frac{1}{2}-2i\bo;z\right)+Bz^{1/2+2i\bo}{_{2}F_{1}}\left(\frac{1}{2}-\bm+2i\bo, 1-\bm,\frac{3}{2}+2i\bo;z\right).
\end{equation}
Having used the formula (\ref{psi_1}), we can come back to the upper component of the spinor wave function $\psi_1$. For the first combination of parameters $\alpha$ and $\beta$, we obtian:
\begin{eqnarray}\label{psi_1_1}
\nonumber\psi_1=(1-z)^{\bm}\left(Az^{i\bo+1/2}{_{2}F_{1}}\left( \bm+\frac{1}{2}+2i\bo, \bm+1,2i\bo+\frac{3}{2};z\right)+\right.\\ \left.Bz^{-i\bo}{_{2}F_{1}}\left(\bm, \bm+\frac{1}{2}-2i\bo,\frac{1}{2}-2i\bo;z\right)\right).
\end{eqnarray}
For the second combination of the parmeters  $\alpha$ and $\beta$, one can arrive at:
\begin{eqnarray}\label{psi_1_2}
\nonumber\psi_1=(1-z)^{-\bm}\left(Az^{i\bo+1/2}{_{2}F_{1}}\left(\frac{1}{2}-\bm+2i\bo,1-\bm, 2i\bo+\frac{3}{2};z\right)+\right.\\ \left.Bz^{-i\bo}{_{2}F_{1}}\left(-\bm,\frac{1}{2}-\bm-2i\bo,\frac{1}{2}-2i\bo;z\right)\right).
\end{eqnarray}
It can be shown that the upper function $\psi_1$ corresponding to the third variant of parameters $\alpha$ and $\beta$ is completely the same as the function (\ref{psi_1_1}). The same fact also takes place for the function (\ref{psi_1_2}) and the upper function that appears for the forth variant of parameters $\alpha$ and $\beta$.  To find a link between the functions (\ref{psi_1_1}) and (\ref{psi_1_2}), a well-known relation for the hypergeometric functions should be used \cite{abramowitz}:
\begin{equation}
{_{2}F_{1}}(a,b,c;z)=(1-z)^{c-a-b}{_{2}F_{1}}(c-a,c-b,c;z).
\end{equation}
Having applied it to the function (\ref{psi_1_2}), we immediately arrive at the conclusion that the upper functions (\ref{psi_1_1}) and (\ref{psi_1_2}) are copletelely the same. So, all the variants for the parametres $\alpha$ and $\beta$ lead to the unique upper fuction $\psi_1$ which can be taken in the form (\ref{psi_1_1}).

To obtain quasinormal modes  the behaviour of the wavefunction (\ref{psi_1_1}) should be analysed at the horizon point and at the infinity. In the vicinity of the horizon point ($z=0$), the wave function $\psi_1$ behaves as 
\begin{equation}
\psi_1\simeq Az^{i\bo+1/2}+Bz^{-i\bo}=Ae^{(i\bo+1/2)\ln z}+Be^{-i\bo\ln z}
\end{equation}
The first term of the given above function would correspond to an outgoing wave solution, whereas the second one gives rise to the ingoing wave. 
According to the quasinormal modes method, it is required that only the ingoing waves exist in the neighbourhood of the horizon.  It leads to the condition that $A=0$. As a result, we arrive at the expression for the upper wave function $\psi_1$:
\begin{equation}
\psi_1(z)=Bz^{-i\bo}(1-z)^{\bm}{_{2}F_{1}}\left(\bm+\frac{1}{2}-2i\bo,\bm,\frac{1}{2}-2i\bo;z\right)
\end{equation} 
To examine behaviour of the wavefunction at the infinity (z=1), a linear transformation $z\rightarrow 1-z$  should be made and Kummer\rq{}s relation for the hypergeometric functions should be used \cite{abramowitz}. Having performed them, we obtain:
\begin{eqnarray}
\nonumber\psi_1(z)=Bz^{-i\bo}(1-z)^{\bm}\left(\frac{\Gamma(-2\bm)\Gamma(1/2-2i\bo)}{\Gamma(-\bm)\Gamma(1/2-\bm-2i\bo)}{_{2}F_{1}}\left(\bm+\frac{1}{2}-2i\bo,\bm,2\bm+1;1-z\right)+\right.\\\left.(1-z)^{-2\bm}\frac{\Gamma(2\bm)\Gamma(1/2-2i\bo)}{\Gamma(\bm)\Gamma(1/2+\bm-2i\bo)}{_{2}F_{1}}\left(-\bm,\frac{1}{2}-\bm-2i\bo,1-2\bm;1-z\right)\right)
\end{eqnarray}
The asymptotic expression for the latter function in the neighbourhood of the infinity takes the form:
\begin{equation}
\psi_1\simeq B(1-z)^{\bm}\frac{\Gamma(-2\bm)\Gamma(1/2-2i\bo)}{\Gamma(-\bm)\Gamma(1/2-\bm-2i\bo)}+B(1-z)^{-\bm}\frac{\Gamma(2\bm)\Gamma(1/2-2i\bo)}{\Gamma(\bm)\Gamma(1/2+\bm-2i\bo)}
\end{equation}
For the upper time dependent componet of the spinor wave function, we obtain:
\begin{equation}
\psi_{up}(z,t)\simeq e^{-i\omega t}\left[B(1-z)^{\bm}\frac{\Gamma(-2\bm)\Gamma(1/2-2i\bo)}{\Gamma(-\bm)\Gamma(1/2-\bm-2i\bo)}+B(1-z)^{-\bm}\frac{\Gamma(2\bm)\Gamma(1/2-2i\bo)}{\Gamma(\bm)\Gamma(1/2+\bm-2i\bo)}\right]
\end{equation}
 Now we should impose the boundary condition  on the spatial infinity. Taking into account the fact that the background geometry (\ref{metric_2}) is not asymptotically flat  and similarly to the previous case  we require that the wavefunction  should vanish at the infinity.  To obey the condition, we should impose that  the argument of the gamma function $\Gamma(1/2+\bm-2i\bo)$ is equal to nonpositive integer: $1/2+\bm-2i\bo=-n$  (the condition which defines the poles of the gamma function). The latter relation allows us to obtain quasinormal frequencies which take the form: 
\begin{equation}\label{qnm_2}
\omega =-i\frac{u_+}{l}\left(n+\bm+\frac{1}{2}\right).
\end{equation}  
and here $n=0,1,2,\ldots$.
The quasinormal frequencies are completely imaginary and the imaginary part is negative for any number $n$. Now we conclude that the black hole metric (\ref{metric_2}) is stable under influence of the fermionic perturbations. To consider  QN modes for the lower component of the wavefunction, we use the relation (\ref{dir_2c}) which can be rewritten in  the form:
\begin{equation}
\psi_2=-\frac{(1-z)}{\bm}\left(\frac{i\bo}{\sqrt{z}}+\sqrt{z}\frac{\partial}{\partial z}\right)\psi_1=-\frac{(1-z)}{\bm}z^{1/2+i\bo}\frac{\partial}{\partial z}z^{i\bo}\psi_1
\end{equation}
As a consequence, the lower component of the wave fuction will be as follows:
\begin{eqnarray}
\psi_2(z)=-\frac{B\bm}{1/2-2i\bo}(1-z)^{\bm}z^{1/2-i\bo}{{_2}F_{1}}\left(\bm+\frac{1}{2}-2i\bo,\bm+1,\frac{3}{2}-2i\bo;z\right)
\end{eqnarray}
In the domain near the horizon point (when $z\rightarrow 0$), the obtained wavefunction will lead to the ingoing wave as it is required. The behaviour of the function $\psi(z)$ should also be investigated at the spatial infinity. To analyse the behaviour at the infinity, we make use of the Kummer\rq{}s transformation again. Making similar analysis as it was performed for the upper component, we can conclude that the lower component would have the same quasinormal frequencies (\ref{qnm_2}) as the upper one. We note that the situation that  for lower components we might have the same (as well as additional different) set of frequencies was  described in the paper \cite{Becar_EPJC}. Comparing the obtained result for quasinormal frequencies with the corresponding result for the scalar perturbation, we can conclude that the black hole is stable under influence of fermionic perturbation of arbitrary mass whereas for the scalar field geometry might be unstable for sufficiently large masses of the field \cite{Cruz_arxiv_2015}. It can be also shown that the imosed boundary conditions at the infinity lead to the vanishing flux, defined by the formula (\ref{flux_1}), so the situation is completely identical to the previous case. 

\section{Concluding remarks}
We studied fermionic quasinormal frequencies for two types of $1+1$ dimensional HL black holes.  The first type of black hole's solution is similar to the corresponding solution, which can be found in the framework of the standard GR.  We impose boundary conditions on the solutions of the Dirac equations to obtain quasinormal modes, namely we demand that in the vicinity of horizon the wavefunction should behave as an ingoing wave and it has to vansih at the infinity because background geometry is not asymptotically flat. The solutions which satisfy mentioned above conditions were found  and correseponding frequencies of the wavefunctions are complex and the imaginary part of the frequencies are bounded from above. The obtained continuous spectrum for the quasinormal modes is in agreement with the results of the work \cite{Estrada-Jimenez_2013}.  We also note that our analysis brings the conclusion that the upper an lower components of Dirac wavefunction have the same spectra. 

The second type of black hole solution is defined in the presence of dilatonic field.  For this type of black hole\rq{}s geometry we impose the same boundary conditions on the solution fo the wave equation. The solution which fulfil the imposed boundary condition have purely imaginary discrete spectrum.  It should be noted that the black hole geometry is stable under the influence of fermionic field of arbitrary mass, and as we mentioned before, for scalar perturbations, black hole might be unstable for large mass of the field \cite{Cruz_arxiv_2015}.

\section{Acknowledgements}
This work was partly supported by Project FF-30F (No. 0116U001539) from the Ministry of Education and Science of Ukraine and Grant No. 0116U005055 
of the State Fund For Fundamental Research of Ukraine.


\begin{thebibliography}{99}
\bibitem{Regge_PR1957} T. Regge, J. A. Wheeler, Phys. Rev. {\bf 108}, 1063 (1957).
\bibitem{Zerilli_PRD1970} F. J. Zerilli, Phys. Rev. D {\bf 2}, 2141 (1970).
\bibitem{Zerilli_PRL1970} F. J. Zerilli, Phys. Rev. Lett. {\bf 24}, 737 (1970).
\bibitem{Nollert_CQG99} H. P. Nollert, Class. Quant. Grav. {\bf 16}, R159 (1999).
\bibitem{Berti_CQG2009} E. Berti, V. Cardoso, A. O. Starinets, Class. Quant. Grav. {\bf 26}, 163001 (2009).
\bibitem{Konoplya_RMP2011} R. A. Konoplya, A. Zhidenko, Rev. Mod. Phys. {\bf 83}, 793 (2011).
\bibitem{aharony_PhysRept2000} O. Aharony, S. S. Gubser, J. Maldacena, H. Ooguri, Y. Oz, Phys. Rept.  {\bf 323}, 183 (2000).
\bibitem{Horowitz} G. T. Horowitz, V. E. Hubeny, Phys. Rev. D {\bf 62}, 024027 (2000). 
\bibitem{Son_JHEP2002} D. T. Son, A. O. Starinets, JHEP {\bf 09}, 042 (2002).
\bibitem{Starinets_PRD02} A. O. Starinets,  Phys. Rev. D {\bf 66}, 124013 (2002).
\bibitem{Nunez_PRD03} A. Nunez, A. O. Starinets,  Phys. Rev. D {\bf 67}, 124013 (2003).
\bibitem{Kovtun_PRD05} P. K. Kovtun, A. O. Starinets,  Phys. Rev. D {\bf 72}, 086009 (2005).
\bibitem{Hod_PRL98} S. Hod, Phys. Rev. Lett. {\bf 81}, 4293 (1998).
\bibitem{Dreyer_PRL2003} O. Dreyer, Phys. Rev. Lett. {\bf 90}, 081301 (2003).
\bibitem{Maggiore_PRL2008} M. Maggiore,  Phys. Rev. Lett. {\bf 100}, 141301 (2008).
\bibitem{Konoplya_PRD2004} R. A. Konoplya, Phys. Rev. D {\bf 70}, 047503  (2004).
\bibitem{Kiefer_CQG2004} C. Kiefer, Class. Quant. Grav. {\bf 21}, L123 (2004).
\bibitem{Horava_PRD09} P. Ho\v{r}ava, Phys. Rev. D {\bf 79}, 084008 (2009).
\bibitem{Horava_JHEP09} P. Ho\v{r}ava, JHEP {\bf 0903}, 020 (2009).
\bibitem{Horava_PRL09} P Horava, Phys. Rev. Lett. {\bf 102}, 161301 (2009).
\bibitem{Kehagias_PLB09} A. Kehagias, K. Sfetsos, Phys. Lett. B {\bf 678}, 123 (2009).
\bibitem{Park_JHEP09} M.-I. Park, JHEP {\bf 09}, 123 (2009).
\bibitem{Cai_PRD09} R. G. Cai, L. M. Cao, N. Ohta, Phys. Rev. D {\bf 80}, 024003  (2009).
\bibitem{Kiritsis_JHEP} E. Kiritsis, G. Kofinas, JHEP {\bf 1001}, 122 (2010). 
\bibitem{Myung_PLB09} Y. S. Myung, Phys. Lett. B {\bf 678}, 127 (2009).
\bibitem{Cai_PLB09} R. G. Cai, L. M. Cao, N. Ohta, Phys. Lett. B {\bf 679}, 504  (2009).
\bibitem{Myung_EPJC10} Y. S. Myung, Y. W. Kim, Eur. Phys. J. C {\bf 68}, 265 (2010).
\bibitem{Koutsoumbas_PRD10} G. Koutsoumbas, P. Pasipoularides, Phys. Rev. D. {\bf 82}, 044046 (2010).
\bibitem{Chen_PLB10} S. Chen, J. Jing, Phys. Lett. B {\bf 687}, 124  (2010).
\bibitem{Konoplya_PLB09} R. A. Konoplya, Phys. Lett. B {\bf 679},499 (2009).
\bibitem{Varghese_MPLA2011} E. Varghese, V. C. Kuriakose, Mod. Phys. Lett. A {\bf 26},1645 (2011).
\bibitem{Varghese_GRG2011} E. Varghese, V. C. Kuriakose, Gen. Rel. Grav.{\bf 43}, 2757 (2011).
\bibitem{Becar_IJMPD13} R. Becar, P. A. Gonzalez, Y. Vasquez, Int. J. Mod. Phys. D {\bf 22}, 1350007  (2013).
\bibitem{Cruz_arxiv_2015} M. Cruz, M. Gonzalez-Espinoza, J. Saavedra, D. Vargas-Arancibia, Eur. Phys. J. C {\bf 76}, 75 (2016).
\bibitem{Becar_EPJC} R. Becar, P. A. Gonzalez, Y. Vasquez, Eur. Phys. J. C {\bf 74}, 2940 (2014).
\bibitem{Bazeia_PRD2015} D. Bazeia, F. A. Brito, F. G. Costa, Phys. Rev. D {\bf 91}, 044026 (2015).
\bibitem{Mann90} R. B. Mann, A. Sheikh, L. Tarasov, Nucl. Phys. B {\bf 341}, 134 (1990).

\bibitem{Estrada-Jimenez_2013} S. Estrada-Jimenez, A. Lopez-Ortega, A. Lopez-Ortega, Gen. Rel. Grav. {\bf 45}, 2239 (2013).
\bibitem{abramowitz} M. Abramowitz, A. Stegun, Handbook of mathematical functions, N.Y., Dover, 1964
\end{thebibliography}
\end{document}